\documentstyle[preprint,aps]{revtex}
\title{Self-consistent theory of superconducting mesoscopic
weak links}
\author{A. Levy Yeyati, A. Mart\'{\i}n-Rodero and F. J. Garc\'{\i}a-Vidal}
\address{
Departamento de F\'\i sica de la Materia Condensada C-XII.\\
Facultad de Ciencias. Universidad Aut\'onoma de Madrid.\\
E-28049 Madrid. Spain.}
\begin{document}

\draft
\maketitle

\begin{abstract}
A microscopic model for describing a superconducting mesoscopic
weak link is presented.
We consider a model geometry consisting of a narrow channel coupled
to wider superconducting electrodes which act as reservoirs fixing the
asymptotic values of the complex order parameter.
For this model, the Bogoliubov-de Gennes
equations are discretized and solved self-consistently using
 a non-equilibrium Green functions formalism.
The transport properties and the electronic excitation spectra of
this system are studied for the different regimes that
can be reached by varying parameters like coherence length,
constriction length, normal transmission coefficient and temperature.
We study in detail the transition
from the point contact limit to the infinite channel length case,
analyzing the maximum Josephson current that can be sustained by the
weak link as a function of its transmission coefficient and length.
It is also shown that for a constriction size ranging from zero to
several times the coherence length, most of the current is
carried, inside the constriction region, by bound states
within the superconducting energy gap. These states correspond
to Cooper pairs with binding energies smaller than the superconducting
gap and which are spatially extended along the channel region,
decaying exponentially inside the reservoirs.
The importance of the self-consistent determination of the order
parameter along the weak link is illustrated by analyzing different
profiles  obtained for channel lengths of the order of
the coherence length. For temperatures not very close to $T_{c}$,
our microscopic calculation predicts the appearance of
features which cannot be obtained from Ginzburg-Landau theory.
\end{abstract}

\vspace{0.3in}

\narrowtext
\section{Introduction}
The interest in the physics of submicron superconducting devices
\cite{devices}
stems to a great extent from the special combination, taking place in
this kind of systems, of normal electron
phase coherence in the nanometer scale together with quantum macroscopic
effects associated with the superconducting state.

Some of the superconducting mesoscopic devices which
are recently receiving more
attention include normal metal-superconductor (NS) junctions and
S-Semiconductor (Sm) junctions where Andreev reflection processes
play a crucial role
\cite{Been2,Lambert,Bagwell,Gunsen},
and superconducting mesoscopic point contacts which can exhibit
quantization of the Josephson current \cite{Been3}.

A theoretical description of the transport properties in all these
submicron devices has to deal with spatial inhomogeneities due to
the presence of geometrical boundaries and interfaces between different
materials;
in a superconducting mesoscopic device this could lead to strong spatial
variations
of the superconducting order parameter.

In normal mesoscopic devices the problem of computing their transport
properties has
been historically addressed by different approaches. One is based on the
scattering
picture first proposed by Landauer \cite{Landauer}
and generalized to more complicated
situations (multi-channel and multi-lead cases) by different authors
\cite{Buttiker}.
Another approach, specially suited to deal with local
inhomogeneities, relies on
the use of a localized representation for the electronic states
of the sample, whose transport
properties can then be efficiently calculated in terms of Green functions
\cite{Green}.

The scattering approach has been extended to superconductors
by Blonder et al. \cite{BTK} and
more recently applied to a mesoscopic NS junction by Beenakker
\cite{Been2}, who has obtained
a multichannel generalization of the Blonder et al. result, and to different
superconducting
devices by other authors \cite{Lambert,Bagwell,Gunsen}. Despite its
many advantages,
the scattering approach cannot describe properly situations in which the
spatial variations
of the superconducting order parameter on a scale smaller
or comparable to the superconducting
coherence length ($\xi_0$) may be important.
For instance, this can be the case in a NS
junction when there is an appreciable induction of superconductivity
in the normal electrode by a proximity effect \cite{Agrait}.
Another example is that
of a superconducting weak link with a length comparable to $\xi_0$,
in which case, as shown in the present paper,
the self-consistent
determination of the order parameter profile becomes
unavoidable.

In a recent letter \cite{Letter}, we have presented a theoretical
approach based on a local description of the superconducting mesoscopic
system, in which a non-equilibrium Green functions formalism is used.
This method provides
an efficient way  of calculating the transport properties and the spatial
variations of the
self-consistent order parameter through the whole system. The aim of the
present
paper is to discuss in further detail our approach, taking as a test system a
mesoscopic
superconducting weak link (SWL) at zero voltage. We shall consider
 a model geometry in which two wide superconducting electrodes are
coupled by a narrow channel. As
discussed by Likharev \cite{Likharev} in his review on weak links,
there are two main reasons for the basic interest on this
kind of systems: first, weak links of reduced dimensions exhibit Josephson
effect in non-tunnel conditions and are  specially  suitable for a variety
of applications. On the other hand, for weak links of increasing length (larger
 separation between electrodes),
 the transition from the Josephson effect to bulk transport
in superconductors can be studied. This last question is addressed in detail in
the
present paper.

Traditionally, the transport properties of SWL have been analyzed with the help
of phenomenological Ginzburg-Landau (GL) theory \cite{Aslamazov}.
However, GL theory is only valid for a limited temperature range
($T \rightarrow T_c$) and, furthermore, the implicit
hypothesis of
slow spatial variation of the order parameter in GL theory as derived
from the microscopic theory \cite{Gorkov},
breaks down for a general mesoscopic geometry.
In these situations a complete self-consistent solution of the
Bogoliubov-de Gennes equations would be necessary \cite{Eilenberger}.
As we show in this work, features in the order
parameter profile associated with length scales smaller than $\xi_0$,
like the Fermi wavelength and some geometrical length scales, emerge in a
natural way from the microscopic calculation. In addition, another
result arising from the microscopic calculation, which could never be
obtained from GL
theory, is the existence of bound states inside the superconducting gap playing
a fundamental role in the transport through the weak link.

The plan of the present paper is the following: in section II we introduce our
discretized
(tight-binding) model for a general constriction geometry, and discuss the
conditions for the fulfillment of current conservation within our model. In
section III the
non-equilibrium Green functions formalism
in a superconducting broken-symmetry
representation is presented, giving the expression for the relevant
quantities within this
formalism and details about its self-consistent determination.
Section IV is devoted to an analysis of the transition from the point
contact regime to infinite 1D-superconductivity,
discussing the maximum current
that can flow as a function of the weak link transmission coefficient.
The analysis of the spectral densities reveals the presence of a slowly
increasing
number of bound states inside the superconducting gap. It is shown that these
states
give the main contribution to the supercurrent in the constriction region and
decay
exponentially inside the 3D electrodes.

 In section V the importance of the
self-consistent determination of the order parameter is illustrated by
discussing the
different type of profiles that can be obtained for a mesoscopic SWL. We
emphasize those
new features that cannot appear in a GL calculation. Finally, in section VI the
effects of
temperature both in the current-phase relationship and self-consistent profiles
are analyzed.
We find that, except very near $T_{c}$, there is no simple scaling of the
system properties
with temperature.

The paper is closed with some concluding remarks.

\section{Discretized model for a mesoscopic weak link}

We consider a model weak link like the one depicted in Fig. 1.
 The constriction width, W,
is assumed to be smaller than the penetration length and comparable to
the Fermi wavelength, $\lambda_{F}$, leading to a small number of conducting
channels.
The constriction length, $L_{c}$, can be varied from the point contact
regime, $L_{c}/\xi_0 \rightarrow 0$,
to the opposite case, $L_{c}/\xi_0 \rightarrow \infty$, recovering in
this case the limit of a homogeneous quasi-1D superconductor. Phase
coherence is assumed to be preserved along the whole system.

The wider regions representing the left and right electrodes ensure the
asymptotic
convergence  of the complex order parameter, $\Delta$, to its bulk value.
This model geometry would be
similar to the ODSEE model proposed by Likharev \cite{Likharev}.

Our aim will be the complete self-consistent solution of the microscopic
Bogoliubov-de Gennes
equations  \cite{de Gennes} for this system. For this purpose we find
convenient to
formulate these equations using a site representation for the electronic
states. This
representation can be viewed either as a tight-binding  description of the
electronic states or as a discretization of the
Bogoliubov-de Gennes equations
(a simple scaling of the parameters would account for the use of one or
another
description). For the electron-electron interaction we make the usual
simplifying
assumption of taking it as a contact attractive interaction
\cite{de Gennes}, which in a
site representation adopts the form of a negative-U Hubbard-like
local term.

Within these assumptions the mean-field model hamiltonian giving rise
to the
Bogoliubov - de Gennes equations for our system can be written as
\cite{Letter}:

\begin{eqnarray}
\hat{H} & = & \sum_{i,\sigma} (\epsilon_i - \mu)
c^{\dagger}_{i \sigma} c_{i \sigma}
+ \sum_{i \neq j, \sigma} t_{ij} c^{\dagger}_{i \sigma} c_{j \sigma}
\nonumber \\
& & + \sum_{i}( \Delta^{\ast}_{i} c^{\dagger}_{i \downarrow}
c^{\dagger}_{i \uparrow} + \Delta_{i} c_{i \uparrow} c_{i \downarrow}),
\end{eqnarray}
\noindent
where, for the zero voltage case, the chemical potential $\mu$
is a constant
throughout the whole system and the sum over $i$ and $j$ in the
second term
is restricted to nearest neighbours only. The self-consistent
conditions for the order parameter on each site are given by:

\begin{equation}
\Delta_{i} = -U_i <c^{\dagger}_{i \downarrow} c^{\dagger}_{i \uparrow}> ,
\end{equation}

\noindent
where $U_i$ is the attractive e-e interaction at site $i$. By choosing
appropriately the $U_{i}$, $t_{ij}$ and $\epsilon_{i}$ in the three
regions (left electrode, constriction,
right electrode) one can model different situations: S-S'-S, S-N-S,
S-Sm-S, etc. In this paper we shall concentrate in the first situation,
assuming
the same $bulk$ value of the order parameter modulus, $|\Delta|$,
on the three separate regions
and fixing the value $\epsilon_i = \mu = 0$ (which implies e-h symmetry)
in the whole system for simplicity.
This criterion fixes the ratio $U_i/t_{ij}$ inside the three regions.
The total phase drop along the whole system $\phi = \phi_L - \phi_R$
is imposed as a boundary condition, $\phi_L$ and $\phi_R$ being the
the bulk value of the order parameter phase on the left and right
electrodes respectively. A relevant parameter is the superconducting
coherence length in the constriction $\xi_0(T)$, which in our model
can be estimated as $\xi_0(T) = 2 t_c/\pi \Delta(T)$, where $t_c$
is the hopping parameter in this region.

The current between two neighbouring sites $ij$ is given by:

\begin{equation}
I_{ij}(t) = \frac{i e}{\hbar} \sum_{\sigma} \left( t_{ij}
< c^{\dagger}_{i\sigma}(t) c_{j \sigma}(t) > - t_{ji}
< c^{\dagger}_{j\sigma}(t) c_{i \sigma}(t) > \right) .
\end{equation}

In the
zero voltage case the supercurrent does not depend explicitly on time.
It is worth noticing that current conservation is only fulfilled when the
solution
of the mean-field superconducting hamiltonian
is fully self-consistent \cite{BTK,Letter,jap,Sols}.
Indeed, current conservation provides a stringent test of
self-consistency.

The proof of this statement
is straightforward when using a site representation.
Starting from the equation of motion for the electron density operator
at site $i$
($\hat{\rho}_{i}=\sum_\sigma c^{\dagger}_{i \sigma} c_{i \sigma}$)
we find:

\begin{eqnarray}
\frac{\partial < \hat{\rho}_{i} >}{\partial t}& = & \frac{i}{\hbar}
< [ \hat{\rho}_{i},\hat{H} ] > \nonumber \\ &  = & - \sum_{j} I_{ij}
+\frac{2ie}{\hbar}(
\Delta_{i}< c_{i \uparrow} c_{i \downarrow}>
-  \Delta^{\ast}_{i} <c^{\dagger}_{i \downarrow}c^{\dagger}_{i \uparrow}>)
\end{eqnarray}

The last term appears due to the fact that $\hat{H}$ does not
conserve the particle
number. However, when the
self-consistency condition (Eq. (2)) is fulfilled, this
term vanishes, and the continuity equation,
$\frac{\partial < \hat{\rho}_{i} >}{\partial t} +
 \sum_{j} I_{ij} = 0$, is recovered for every site .

\section{Solution in terms of non-equilibrium Green functions}

The averaged quantities appearing in Eqs. (2) and (3) can be expressed in terms
of non-equilibrium Green functions \cite{Keldysh}. For
the description of the superconducting state it is useful to introduce spinor
field operators \cite{Nambu}, which in a site representation are defined as:

\begin{equation}
\hat{\psi_{i}} = \left(
\begin{array}{c}
c_{i \uparrow} \\ c^{\dagger}_{i \downarrow}
\end{array} \right) \hbox{  ,     } \hat{\psi}^{\dagger}_{i}=
\left(
\begin{array}{cc}
 c^{\dagger}_{i \uparrow} & c_{i \downarrow}
\end{array} \right)
\end{equation}
\noindent
Then, the different correlation functions appearing in the Keldysh formalism
adopt the standard causal form:

\begin{equation}
\hat{G}_{ij}^{\alpha,\beta}(t_{\alpha},t'_{\beta})=-i
< \hat{T}[\hat{\psi}_{i}(t_{\alpha})
\hat{\psi}_{i}^{\dagger}(t'_{\beta})]>
\end{equation}
\noindent
where $\hat{T}$ is the chronological ordering operator along the closed time
loop
contour \cite{Keldysh}. The labels $\alpha$ and $\beta$ refer to the upper
($\alpha \equiv +$)
and lower ($\alpha \equiv -$) branches on this contour. The correlation
functions
$\hat{G}^{+-}_{ij}$, which can be associated
within this formalism with
the electronic non-equilibrium distribution functions \cite{Kadanoff},
are given by the (2x2) matrix:

\begin{equation}
{\bf G}^{+-}_{i,j}(t,t^{\prime})= i \left(
\begin{array}{cc}
<c^{\dagger}_{j \uparrow}(t^{\prime}) c_{i \uparrow}(t)>   &
<c_{j \downarrow}(t^{\prime}) c_{i \uparrow}(t)>  \\
<c^{\dagger}_{j \uparrow}(t^{\prime}) c^{\dagger}_{i \downarrow}(t)>  &
<c_{j \downarrow}(t^{\prime}) c^{\dagger}_{i \downarrow}(t)>
\end{array}  \right) .
\end{equation}
\noindent
Eqs. (2) and (3) can then be written in terms of the
 Fourier transform matrix elements of
$\hat{G}^{+-}_{ij}(t,t')$:

\begin{equation}
\Delta_i = -\frac{|U_i|}{2 \pi i} \int^{\infty}_{-\infty} d\omega [{\bf G}
^{+-}_{ii}(\omega)]_{21},
\end{equation}

\begin{equation}
I_{ij} =\frac{2 e}{h}  \int^{\infty}_{-\infty} d\omega \left( t_{ij}[{\bf
G}^{+-}_{j,i}
(\omega)]_{11} -  t_{ji} [{\bf G}^{+-}_{i,j}(\omega)]_{11} \right).
\end{equation}
\noindent

For the zero voltage case the calculation of the different
$\hat{G}^{+-}(\omega)$ is particularly simple because the following
relation holds:

\begin{equation}
\hat{G}^{+-}_{ij}(\omega)=f(\omega)[\hat{G}_{ij}^{a}(\omega)-\hat{G}_{ij}^{r}(\omega)]
\end{equation}

\noindent
where $f(\omega)$ is the Fermi distribution function, and $\hat{G}^{a,r}$ are
the
advanced and retarded Green functions,
which can be computed using recursive techniques \cite{Yeyati}.
(Note that this relation is the same as in a currentless state).

The Green functions must be calculated
self-consistently, according to Eq.(8).
This is achieved starting from an initial guess for the order parameter
profile and then using an iterative algorithm. As reported in
\cite{Letter}, the electrodes can be modelled in a simple way by Bethe
lattices. This choice both facilitates the computation of the Green
functions and ensures a fast spatial convergence of the order parameter
to its bulk values. The Bethe lattice geometry is able to simulate
in a simple way
the $geometrical \;\ dilution$ of the current taking place when passing
from a quasi-1D to a 3D structure. We have fixed the Bethe lattice
coordination number $z + 1 = 4$, which leads to a convergence of the
order parameter within three or four layers.

\section{Transition from point contact to 1D-flow}

It is interesting to analyze in detail the maximum d.c. Josephson
current, $I_c$ that can be sustained by a SWL
with a single conducting channel as a function of its length and
transmission coefficient, $\alpha$. Within our tight-binding model
$\alpha$ is a known function of the hopping parameters $t_{ij}$. For the
results presented in this work $\alpha$ is varied by changing uniformly
$t_{ij}$ inside the constriction, while keeping them fixed inside
the reservoirs.

For any value of $\alpha$,
the point contact limit ($L_c/\xi_0 \rightarrow 0$) is well understood
\cite{Been3,Letter,Arnold}. In this case, the phase profile can be
well approximated by a step function and the current-phase relationship
can be obtained analytically for a symmetric junction
(details of this derivation within our
formalism are given in Appendix A):

\begin{equation}
I(\phi) = \frac{e \alpha}{2 \hbar}
\frac{|\Delta(T)|^{2}}{|\epsilon(\phi)|} \sin(\phi)
\tanh[\frac{|\epsilon(\phi)|}{2 k_B T}] ,
\end{equation}

\noindent
where $\epsilon(\phi) = \pm|\Delta(T)|\sqrt{1 - \alpha\sin^2(\phi/2)}$
denotes the position of bound states within the superconducting gap.
These bound states give the main contribution to the supercurrent
across the interface. As we shall see the existence of bound states
inside the gap remains even for a SWL of length much larger than $\xi_0$.
It should be mentioned that an expression essentially equivalent to
Eq. (11) was first derived by Kulik and Omel'yanchuk within a
semiclassical approximation to the Bogoliubov - de Gennes equations
\cite{Kulik}

The maximum supercurrent in this point contact limit can be obtained
as a function of $\alpha$ from Eq. (11). At zero temperature we find:

\begin{equation}
I_{c,0}(\alpha)=\frac{e\Delta}{\hbar}(1-\sqrt{1-\alpha})
\end{equation}
\noindent

Therefore the maximum possible current through a single quantum
channel is
$e\Delta/\hbar$ whereas in the tunnel regime
$I_{c,0} \simeq e\Delta\alpha/2\hbar$, which corresponds
to the well known Ambegaokar-Baratoff value $\pi \Delta/2eR_{N}$
\cite{Ambegaokar} for the single channel case.
The function $I_{c,0}(\alpha)$ is plotted in Fig. 2.
Let us mention that the divergence in the
derivative $\partial I_{c,0}/\partial \alpha$ as $\alpha$ tends to 1 is
a zero temperature feature, which is quickly smoothed at finite
temperatures;
at $T=0.2 T_{c}$, $I_{c,0}(\alpha=1)$ is reduced by a factor of
$\sim 20 \%$.
This suggests that the experimental observation of the quantized
critical current
$e\Delta/\hbar$ requires quite delicate conditions (i.e., perfect
transmission, symmetric junction, very low temperatures, etc).

At zero temperature,
when $L_{c}$ is increased keeping the value of $\alpha$ fixed, the maximum
current can either decrease or increase from the $L_{c} = 0$ value.
In Fig. 3 we represent
the behaviour of $I(\phi)$ with increasing $L_{c}$ for three values of the
transmission $\alpha$. Only the positive part of $I(\phi)$ is plotted.
A general trend is the appearance of multivaluation for $L_c/\xi_0
> 1$, although the precise value for this threshold ratio
is dependent on $\alpha$, as can be observed in Fig 3.
Note that the form of the $I(\phi)$ curves itself is
strongly dependent on $\alpha$; therefore there is not a simple scaling
between the different set of curves. For instance, the low transmission
case (Fig. 3c) becomes markedly different from the other two when
$L_c/\xi_0$ increases (note the peculiar form of the 48-sites case in
which there appears a cusp at $\phi=2\pi$; for larger lengths this cusp
progressively bends down while $I(\phi=2\pi)$ tends to zero).
In the limit $L_c/\xi_0 \rightarrow \infty$, the behaviour would be
that of a double tunnel junctions with a $I(\phi) \sim \sin(\phi/2)$
characteristic for the upper branch \cite{Letter}.

The maximum current $I_c$ is represented in Fig. 4
as a function of $L_c/\xi_0$ for the three cases shown is Fig. 3 .
It can be observed that when $L_c/\xi_0 \gg 1$,
and for high transmission
$I_{c}$ decreases asymptotically to a limiting value, whereas for low
transmission the trend is the opposite. On the other hand, when
$L_c/\xi_0 \sim 1$ interference effects can lead to a non-monotonic
behaviour of $I_c$ with length. See for instance the intermediate
transmission case ($\alpha = .75$) where a dip at $L_c \sim \xi_0$
is found.

The limiting case $L_{c}/\xi_{0} \rightarrow \infty$
corresponds to a homogeneous flow in a one dimensional superconductor
in which the coupling to the
left and right electrodes acts as boundary conditions.
The asymptotic value of the maximum current $I_{c,\infty}$ depends on
this coupling and
is therefore a function of $\alpha$. These values are
represented by the triangles in Fig. 2, where they can be compared with
the $L_{c}=0$ case. It is worth noticing that for perfect transmission
$I_{c,\infty}$ coincides
with the depairing current of a one dimensional superconductor, i.e.
$2 e \Delta/\pi \hbar$ \cite{Bagwell2}. The derivation
of this result within our model is given in Appendix B.
The self-consistent phase gradient
along the linear chain is equal to $2\Delta/\hbar v_{F}$
when this limit is reached.

On the other hand, for low transmission, when the matching to
the electrodes is poor, the
self-consistent phase drop concentrates mainly on the contacts
and, as mentioned above, the system becomes
equivalent to a double tunnel junction. In these conditions
the maximum current is
controlled basically by the transmission through a single junction.
For a symmetric weak
link this single junction transmission $\alpha_{1}$,
is related to the total transmission $\alpha$ by
$\alpha_{1} \approx 2 \sqrt \alpha$
(this relation holds only for $\alpha \ll 1$).
Then, according to Eq. (11) $I_{c,\infty}(\alpha) \approx
\frac{e\Delta}{\hbar} \sqrt{\alpha}$, which is qualitatively
in agreement with
the numerical results for small $\alpha$ in Fig. 2.

A deeper insight on the transition from the point contact to the
infinite 1D case can be
obtained by analyzing the evolution of the local
quasi-particle spectral density,
$\rho_{i}(\omega)$ and the associated current density $j_{i}(\omega)$,
given by:

\begin{equation}
\rho_{i}(\omega) =\frac{1}{\pi} Im [\hat{G}^{a}_{ii}(\omega)]_{11}
\end{equation}

\begin{equation}
j_i (\omega) =\frac{2 e}{h}   \left( t_{i,i+1}[{\bf G}^{+-}_{i+1,i}
(\omega)]_{11} -  t_{i+1,i} [{\bf G}^{+-}_{i,i+1}(\omega)]_{11} \right)
\end{equation}

\noindent
where site $i$ is chosen to be located at the center
of the constriction, where
the effect of the 3D reservoirs is less pronounced.
In the following discussion only energies
$\omega \leq 0$ are considered as $\rho_i(\omega) = \rho_i(-\omega)$
due to e-h symmetry.

For $L_{c}/\xi_{0} \ll 1$ the most relevant feature is the
appearance of a bound
state in the spectra inside the gap. The position and weight of
this state is a function of
the phase difference, $\phi$, and the transmission coefficient, $\alpha$,
as noted in the discussion of Eq. (11). In
this regime this state gives the main contribution to the current
in the constriction:
$j_{i}(\omega)$ is essentially a delta function at the bound state energy.
In Fig. 5 the evolution
of $\rho_{i}(\omega)$ and $j_{i}(\omega)$ as $L_{c}/\xi_{0}$ increases
is represented for
the case of perfect transmission. In all the cases plotted in this
figure the current is the
maximum one for the corresponding constriction length.

Two main features are observable in these curves: first, with increasing
$L_{c}$ new bound
states appear in the gap. The new states initially split from the
continuum, moving to energies closer to $\mu=0$.
Eventually, for $L_{c} \rightarrow \infty$ these states would
fill the gap as shown in the uppermost curves of Fig. 5, which correspond
to the uniform infinite 1D superconductor carrying the critical
current (see Appendix B).
The number of bound states increases very slowly with $L_{c}/\xi_{0}$.
For instance, for $L_{c}/\xi_{0}$ as large as
10, only three bound states are present.

On the other hand, the current density of the
continuous part of the spectrum ($\omega \leq -\Delta$), which at
$L_c/\xi_0 \rightarrow 0$ is negligible, becomes
more important with increasing length, and in the
limit $L_c/\xi_0 \rightarrow \infty$
gives a contribution which has the opposite sign to the
total current.

To conclude this section,
let us analyze briefly the nature of the bound states.
The study of their spatial distribution shows that they correspond to
Cooper pairs that are extended along the constriction region and decay
exponentially inside the reservoirs within a typical length $\sim
\xi_0$. In Fig. 6 we represent the local
spectral weight of the bound states along the
constriction. It corresponds to case (d) in Fig. 5, in which three bound
states are present. As can be seen in Fig. 6, the bound state closer
to $\omega = 0$ has always a nodeless form, while the number of
oscillations increases for the bound states with a larger binding
energy. This situation is reminiscent of the one found for a potential
well; however, one should keep in mind that the bound states in the
present case are due to variations of the superconducting phase
on a finite spatial region (the constriction region), disappearing
when the total phase drop is zero.
In junctions of the type
S-S'-S or S-N-S with large variations in the pairing potential
one could find bound states even without current
\cite{Gunsen,de Gennes,Arnold2}.
We should comment that the spatial distribution
of the current carried by a bound state follows closely
that of its weight. Actually, we have verified that the ratio
between current and weight for each bound state is practically
a constant along the constriction region. This result further
illustrates
the fact that the bound states correspond to Cooper pairs
with a well defined velocity inside the constriction.

\section{Self-consistent order parameter profiles at zero temperature}

The self-consistent determination of the order parameter profiles is essential
for a weak
link of length $L_{c} \geq \xi_{0}$, whereas for $L_{c} \ll \xi_{0}$, the point
contact limit, the detailed form of the self-consistent profiles becomes
irrelevant. In
this latter case the current-phase
relationship is given to a high degree of accuracy by Eq. (11)
\cite{Been3,Letter}.

In this section we present results for $L_{c} \geq \xi_{0}$ in order to
illustrate the
effects of self-consistency. Although some of the overall features
appearing in
the profiles
can be expected on intuitive physical grounds, their
detailed form reflects a
complex interplay between the different model parameters.

In Fig. 7 we represent the phase and modulus profiles for fixed
$L_{c}$ and $
\phi$, and three different values of the transmission coefficient.
All three results
correspond to the upper branch of the $I(\phi)$ characteristic.

The common general features of the profiles along the upper branch are
displayed in these
figures. They consist on a constant phase gradient along the constriction
together with
localized drops at the contacts with the reservoirs. One could draw an
analogy between these
localized phase drops and the voltage drops at the contacts with the leads
in a normal mesoscopic sample (Sharvin resistance) \cite{Buttiker,Pablo}.
On the other hand, the modulus is on average constant along the
constriction.

Superimposed to this general structure, oscillations with a spatial period
$\lambda_{F}/2$
can be observed \cite{Letter}. These oscillations have a maximum
amplitude in the proximity of
the contacts and decay when moving inside
the constriction with a typical decaying length $\sim
\xi_{0}$. This behaviour is most clearly seen
 in case (c) of Fig. 7, which corresponds
to $L_{c}/\xi_{0} \gg 1$ and low transmission,
leading to
oscillations which are more concentrated near the
contacts, having a larger amplitude due to
 the presence of larger ``barriers".
This interference phenomenon is a
consequence of the phase coherence
of normal electrons along the mesoscopic system.
As it is shown in the next section, there is a
gradual disappearance of
the oscillations with increasing temperature.
Let us remark that a similar oscillation
pattern can be found for the electrostatic
potential along a normal mesoscopic
constriction \cite{Pablo,Butt2,Pasta}.

When $I(\phi)$ is a multivalued function,
the solutions in the lower branch exhibit a very
different character: they correspond
to the solitonic-like solutions predicted by GL theory
\cite{Likharev,Sols,Langer}. In Fig. 8 we show
the phase and modulus profiles for this second
type of solution, corresponding to three
different values of $L_{c}$ with fixed $\alpha$ and
$\phi$. We have chosen $ \phi \sim \pi$
where the specific features of these
solutions are more pronounced. It can be observed
that the gross features are the ones
predicted by GL theory, namely an abrupt phase drop
together with a significant depression
of the modulus which nearly goes to zero
at the center of the constriction. Note that this
profile leads to a very low value of the current.
In addition to this general
shape, an oscillation pattern of period
$\lambda_{F}/2$ is again found. An
interesting feature is the existence of a well defined
``core" region of length $\sim \xi_{0}$
where the phase drop takes place and the modulus is
nearly zero (see Fig. 8). This core
remains practically
unchanged with increasing $L_{c}$. This is a feature that
cannot be predicted by GL
theory, the reason being, as we shall see in the next section,
that the core size vanishes when
$T$ approaches $T_{c}$.

The question about the stability of these
type of solution deserves some attention. As
discussed by Langer and Ambegaokar \cite{Langer},
these solutions correspond for a homogeneous
infinite system to saddle points of the free energy.
However, for a finite weak link these
solutions are presumably unstable \cite{Likharev}.
As a matter of fact, we have verified that
these solutions cannot be reached using a simple
iterative algorithm no matter how close from the
actual solution one
starts. Instead, the use of a more sophisticated
algorithm (Broyden type \cite{Vander}) capable of
obtaining any sort of extrema, leads to the
solitonic solution without difficulties.

\section{Effects of finite temperature}

In this section we discuss the effects of
a finite temperature on the $I(\phi)$
characteristic and on the self-consistent
profiles for our model weak link. Although
$L_{c}/\xi_{0}(T)$ is the main parameter
controlling the different regimes (point contact,
infinite homogeneous superconductor)
there is not a simple scaling of the system properties
as a function of this single parameter.

This becomes apparent when studying
the evolution of the different properties with
temperature. Fig. 9 shows the current-phase
relationship for the same set of parameters
as in Fig. 2b and three increasing values
of temperature: 0.3, 0.5 and 0.95
$T_{c}$. As one can observe, the phase interval
where $I(\phi)$ is multivalued gradually
disappears while
$I(\phi)$ tends
to the familiar $sin \phi$ dependence as $T$
$\rightarrow$ $T_{c}$. This behaviour can be
qualitatively understood noticing that for a
given length and $T$ sufficiently high
($ T \rightarrow$ $T_{c}$), the condition
$L_{c}/\xi_{0}(T) \ll 1$ is fulfilled, and
the system should behave like a point contact in which case
Eq. (11) approximately holds.
Note, however, in Fig. 9c, that a significant deviation
from the maximum current value predicted by Eq. (11),  i.e
$I_{c}=\frac{\pi}{2eR_{N}}\frac{\Delta^{2}(T)}{K_{B}T}$
is observed as soon $L_{c}$ is a
small fraction of $\xi_{0}(T)$. This is due to the
fact that the self-consistent phase profile, even for
small $L_c / \xi_0 (T)$, deviates from a step function form.

The possible scaling of the maximum
supercurrent with $L_c / \xi_0 (T)$ can be
analyzed with the help of Fig. 10. For these
values of the transmission coefficient ($\alpha \sim 0.75$),
a universal behaviour of the type $\sim \exp^{-A L_c / \xi_0 (T)}$
is only observed for temperatures larger than
$0.5T_c$. For lower transmission this departure from a universal
behaviour is even more pronounced, the maximum supercurrent  at low
temperatures is in
this case an increasing function of $L_c / \xi_0 (T)$ (see bottom
curve in Fig. 4) while
the $ \exp^{-A L_c / \xi_0 (T)}$ behaviour is only recovered
for $T \rightarrow T_{c}$.

Let us briefly comment the effects of finite
temperature on the self-consistent order
parameter profiles.
Either in the upper and lower branch
solutions there is a gradual disappearance of
the $\lambda_{F}/2$ oscillations (see Fig. 11). In
the upper branch solutions there are
no notable changes in the phase profiles, besides the
smoothing of the oscillations pattern, the
expected Josephson current decrease being due to
the global lowering of $\Delta (T)$ in the whole system.
The evolution of the solitonic
profiles with temperature is more unusual.
This is illustrated in Fig. 11, where it can be
observed how the solitonic ``core", which at
zero temperature has a size $\xi_{0}(0)$,
shrinks and eventually disappears as
temperature increases.
This evolution is particularly clear
in the phase profiles of Fig. 11.

\section{Concluding remarks}

We have presented a self-consistent solution of the microscopic
Bogoliubov-de Gennes equations for a mesoscopic SWL.
As illustrated in the present work, our method allows us to
analyze situations where spatial variations of the superconducting
order parameter over length scales smaller or comparable to the
coherence length are important, and
its self-consistent determination becomes necessary.
As discussed in this work, the usual GL approach
may not be valid for a mesoscopic sample at low temperatures.
Our results reveal that for low temperatures and due to coherence
effects, there is no universal
behaviour of the SWL properties as a function of $L_c/\xi_0(T)$.
These coherence effects are also reflected in the self-consistent
order parameter profiles, which exhibit features
that cannot be predicted by GL theory.

We have analyzed in detail the transition from the point contact
limit to the infinite 1D superconductor characterized by a
continuous spectrum.
The appearance of bound states seems to be a general
feature of a weak link of mesoscopic size due to the spatial
inhomogeneity of the pairing potential in the constriction region.
Even in the absence of a well defined pairing potential well,
spatial variations of the superconducting phase associated with
a supercurrent can lead to the formation of bound states.

The application of the present approach to the description
of submicron superconducting devices of current
experimental interest \cite{devices}, which may include S-N or S-Sm interfaces
and where the sample geometry plays an important role,
is under progress in our laboratory.

\acknowledgements
Support by Spanish CICYT (Contract No. PB89-0165) is acknowledged.
The authors are indebted to F. Flores, J. C. Cuevas and F. Sols for useful
discussions.

\appendix
\section{}

In this appendix we give a short account on the derivation
of the supercurrent-phase relationship for a SWL in the point contact
limit. Starting from Eq. (9), the current evaluated at the interface
between the electrodes can be written as:

\begin{equation}
I =\frac{2 e}{h}  \int^{\infty}_{-\infty} d\omega f(\omega)
\, Tr \left[ {\bf T}_{LR}({\bf G}^a_{RL} - {\bf G}^r_{RL})
- {\bf T}_{RL}({\bf G}^a_{LR} - {\bf G}^r_{LR}) \right]_{11}
\end{equation}

\noindent
where $L$ and $R$ indicate the left and right electrode and $Tr$ denotes
trace over orbitals at the interface. The hopping elements in the
superconducting broken-symmetry representation are given by
$ 2 \times 2$ matrix ${\bf T}_{i,j} = {\bf \tau}_3  t_{i,j}$, where
${\bf \tau}_3$ is a Pauli matrix and $i, j$ denote any pair of orbitals.

The Green functions appearing in this expression can be obtained from
the ones corresponding to the uncoupled electrodes (${\bf T}_{LR} =
0$) ${\bf g}^{a,r}_L$, ${\bf g}^{a,r}_R$, using Dyson equations:

\begin{equation}
{\bf G}^{a}_{LR} = {\bf g}^a_{L}{\bf T}_{LR}{\bf g}^a_{R}{\bf D}^a_{LR}
\end{equation}

\noindent
where ${\bf D}^a_{LR} = \left[ {\bf I} - {\bf T}_{LR}{\bf g}^a_{R}
{\bf T}_{RL}{\bf g}^a_{L} \right]^{-1}$ with similar expressions for
the retarded quantities. Substituting in Eq. (A1) and performing some
elementary algebra, we obtain:

\begin{equation}
I =\frac{2 e}{h}  \int^{\infty}_{-\infty} d\omega f(\omega)
\, Tr \left[ ({\bf D}^a_{RL} - {\bf D}^a_{LR})
- ({\bf D}^r_{RL} - {\bf D}^r_{LR}) \right]_{11}
\end{equation}

This equation is valid for a general junction geometry with any number
of conducting channels. For the simple case of a symmetric junction
with a single channel, an analytical expression can be derived for
$I(\phi)$ if a step-like phase-profile is assumed:

\begin{equation}
I =\frac{4 ei}{h}|t_{LR}|^2 \sin \phi \int^{\infty}_{-\infty}
d\omega f(\omega) \left[
\frac{\tilde{g}^{a}_{L,21}(\omega) \tilde{g}^{a}_{R,12}(\omega)}
{\det(D^a_{LR}(\omega))} -
\frac{\tilde{g}^{r}_{L,21}(\omega) \tilde{g}^{r}_{R,12}(\omega)}
{\det(D^r_{LR}(\omega))} \right]
\end{equation}

\noindent
where the tilde indicates that the phase factor has been removed,
i.e. $\tilde{g}_{L,21} = \exp^{-i\phi_L} g_{L,21}$ and
$\tilde{g}_{R,12} = \exp^{i\phi_R} g_{R,12}$.
The integration in Eq. (A4) can be performed analytically as a
contour integration by
realizing that the main contribution (up to corrections of the
order $\Delta/\epsilon_F$) is given by the zeroes within the
superconducting gap of $\det(D^a_{LR}(\omega))$, which
correspond to
the bound states commented in section IV. The contribution
from these poles yields Eq. (11) straightforwardly.

\section{}

A simple derivation of the depairing current for a 1D superconductor
within our tight-binding model is given below.

The hamiltonian in this case is the one given in Eq. (1) with $t_{ij}
= t (\delta_{i,j+1} + \delta_{i,j-1})$, $\epsilon_i = 0$ and
$\Delta_j = \Delta_q \exp^{iqj}$. This last relation holds for a
uniform superconducting flow, with Cooper pairs of net momentum $q$
\cite{de Gennes}.  For an infinite system, the Green functions adopt
a simpler form in the $k$-representation:

\begin{equation}
[{\bf G}_{k,q} (\omega)]_{11} = \frac{\omega + \epsilon_{-k+q}}
{(\omega - \epsilon_k)(\omega + \epsilon_{-k+q}) - \Delta_q^2}
\end{equation}

\begin{equation}
[{\bf G}_{k,q} (\omega)]_{12} = -\frac{\Delta_q}
{(\omega - \epsilon_k)(\omega + \epsilon_{-k+q}) - \Delta_q^2}
\end{equation}

\noindent
where $\epsilon_k = 2 t \cos k - \mu$. The poles of these functions
at $E^{\pm}_{k,q} = \delta \epsilon_{k,q} \pm \sqrt{
\bar{\epsilon}_{k,q}^2 + \Delta_q^2}$, where $\delta \epsilon_{k,q}
= (\epsilon_k - \epsilon_{-k+q})/2$ and $\bar{\epsilon}_{k,q} =
(\epsilon_k + \epsilon_{-k+q})/2$, give the excitation spectrum of
a current carrying 1D superconductor.
The critical momentum $q_c$ which corresponds to the
condition that the gap in the excitation spectrum goes to zero,
is given, for sufficiently small $q_c$ ($q_c \sin k_F \ll 1$), by
$q_c = \Delta_0 / t \sin k_F$.

It can be shown that the self-consistent order parameter $\Delta_q$
for $q \le q_c$  is in this limit equal to $\Delta_0$
up to corrections of order $q^2$. Actually, these corrections are
positive, leading to a small increase in $\Delta_q$ with increasing
current.

The local spectral density $\rho(\omega)$ and the local current
density $j(\omega)$ discussed in section IV can be directly calculated
from ${\bf G}_{k,q}(\omega)$. In particular, the total current is given
by:

\begin{equation}
I = \frac{4 e}{\pi h} t \int^{\pi}_{-\pi} dk \sin k
\int^{\infty}_{-\infty} d\omega f(\omega)
Im \left[ {\bf G}_{k,q}^a(\omega) \right]_{11}
\end{equation}

At zero temperature, the integral over $\omega$ can be easily
performed giving:

\begin{equation}
I =\frac{e t}{\pi \hbar} \int_{-\pi}^{\pi} dk
\frac{\bar{\epsilon}_{k,q} \sin k}{\sqrt{\bar{\epsilon}_{k,q}^2 +
\Delta_q^2}} \simeq \frac{2e}{\pi \hbar} t \sin k_F \;\; q
\end{equation}

\noindent
where the last approximation holds in the limit $q \sin k_F \ll 1$.
Thus, for $q = q_c$ the depairing current is simply $\frac{2}{\pi}
e \Delta/\hbar$, in agreement with ref. \cite{Bagwell2}.

\begin{figure}
\caption{Schematic representation of our discretized model weak link.}
\end{figure}

\begin{figure}
\caption{Maximum supercurrent for a single channel weak link at
zero temperature in the point contact limit ($L_c=0$) and in the
asymptotic infinite length limit, as a function of the
transmission coefficient.}
\end{figure}

\begin{figure}
\caption{Supercurrent-phase characteristics at zero temperature
for different constriction
lengths and three pair of values for the transmission and the
coherence length:
(a) $\alpha = 1$ and $\xi_0 = 22.05$; (b) $\alpha = .75$ and
$\xi_0 = 12.73$; (c) $\alpha = .28$ and $\xi_0 = 6.37$. The
coherence length and the $L_c$ values given in the figure are measured
in units of the site spacing.}
\end{figure}

\begin{figure}
\caption{Maximum supercurrent at zero temperature
for the same three cases as in Fig. 3
as a function of $L_c/\xi_0$.}
\end{figure}

\begin{figure}
\caption{Local spectral and supercurrent density at the center of the
constriction for the $\alpha = 1$ case and different values of
$L_c/\xi_0$: (a) $L_c/\xi_0 = 2.18$, (b) $L_c/\xi_0 = 5.80$, (c)
$L_c/\xi_0 = 9.43$, (d) $L_c/\xi_0 = 13.06$ and (e) $L_c/\xi_0
= \infty$. The curves are displaced upwards with respect to case (a)
for clarity. All cases correspond to the maximum supercurrent.}
\end{figure}

\begin{figure}
\caption{Spectral weight along the constriction for the bound
states appearing in case (d) of Fig. 5. The order from top to
bottom corresponds to increasing binding energy.}
\end{figure}

\begin{figure}
\caption{Upper branch
self-consistent order parameter profiles (phase and
modulus) for a total phase drop $\phi = 4.60$ and fixed
constriction length $L_c = 88$. The transmission and coherence
length correspond to the three cases shown in Fig. 3. The
curves (a) and (b) are displaced upwards with respect to case (c).}
\end{figure}

\begin{figure}
\caption{Lower branch
self-consistent order parameter profiles (phase and
modulus) for a total phase drop $\phi = 3.20$, $\alpha = .75$,
$\xi_0 = 12.73$ and three values of the constriction length:
(a) $L_c = 48$, (b) $L_c = 64$ and (c) $L_c = 88$.}
\end{figure}

\begin{figure}
\caption{Supercurrent-phase characteristics for $\alpha = .75$,
different constriction
lengths and three values of temperature:
(a) $T = 0.3 T_c$ and $\xi_0(T) = 12.78$; (b) $T = 0.5 T_c$ and
$\xi_0(T) = 13.37$; (c) $T = 0.95 T_c$ and $\xi_0(T) = 39.79$.}
\end{figure}

\begin{figure}
\caption{Maximum supercurrent for $\alpha = .75$
as a function of $L_c/\xi_0(T)$ for increasing values of
temperature.}
\end{figure}

\begin{figure}
\caption{Lower branch
self-consistent order parameter profiles (phase and
modulus) for a total phase drop $\phi = 3.20$, $\alpha = .75$,
$L_c = 88$ and three values of temperature:
(a) $T = 0$, (b) $T = 0.3 T_c$ and (c) $T = 0.5 T_c$.}
\end{figure}


\begin{references}
\bibitem{devices}
A. Kastalsky, A. W. Kleinsasser, L. H. Greene, R.
Bhat, F. P. Milliken and J. P. Harvison, Phys. Rev. Lett. {\bf 67},
3026 (1991);
C. Nguyen, H. Kroemer and E. L. Hu, Phys. Rev. Lett.
{\bf 69}, 2847 (1992);
U. T. Petrashov, V. N. Antonov, P. Delsing and T. Claeson, Phys. Rev.
Lett. {\bf 70}, 347 (1993);
P. Xiong, G. Xiao and R. B. Laibowitz, Phys. Rev. Lett. {\bf 71}, 1907
(1993).
\bibitem{Been2} C. W. J. Beenakker, Phys. Rev. B {\bf 46}, 12841 (1992).
\bibitem{Lambert} C.J. Lambert, V.C. Hui and S.J. Robinson,
J. Phys.: Condens. Matter {\bf 5}, 4187 (1993).
\bibitem{Bagwell} P. F. Bagwell, Phys. Rev. B {\bf 46}, 12573 (1992).
\bibitem{Gunsen} U. Gunsenheimer, U. Schussler and R. Kummel,
Phys. Rev. B {\bf 49}, 611 (1994).
\bibitem{Been3} C. W. J. Beenakker and H. van Houten, Phys. Rev. Lett.
{\bf 66}, 3056 (1991); C. W. J. Beenakker, Proc. 14th. Taniguchi Int.
Simp. on ``Transport Phenomena in Mesoscopic Systemsoo
ed. by  H. Fukuyama and T. Ando (Springer, Berlin, 1992).
\bibitem{Landauer} R. Landauer, Phil. Mag. {\bf 21}, 863 (1970).
\bibitem{Buttiker} M. Buttiker, Y. Imry, R. Landauer and S. Pinhaes,
Phys. Rev. {\bf 31}, 6207 (1985);
M. Buttiker, Phys. Rev. Lett. {\bf 57}, 1761 (1986).
\bibitem{Green} A. D. Stone, Phys. Rev. Lett. {\bf 54}, 2692 (1985);
 J. Ferrer, A. Mart\'{\i}n-Rodero and F. Flores, Phys. Rev. B {\bf 38},
10113 (1988); J. Masek and B. Kramer, Z. Phys. B {\bf 75}, 57 (1989);
A. Levy Yeyati, J. Phys.: Condens. Matter {\bf 2}, 6533 (1990).
\bibitem{BTK} G. E. Blonder, M. Tinkham and T. M. Klapwijk, Phys. Rev.
B {\bf 25}, 4515 (1982).
\bibitem{Agrait} N. Agrait, J. G. Rodrigo, C. Sirvent and
S. Vieira, Phys. Rev. B {\bf 48}, 8499 (1993).
\bibitem{Letter} A. Mart\'{\i}n-Rodero, F. J. Garc\'{\i}a-Vidal and
A. Levy Yeyati, Phys. Rev. Lett. {\bf 72}, 554 (1994).
\bibitem{Likharev} K. K. Likharev, Rev.Mod.Phys. {\bf 51}, 101 (1979).
\bibitem{Aslamazov} L. G. Aslamazov and A. I. Larkin, Zh. Eksp. Teor.
Fiz. Pis'ma Red. {\bf 9}, 150 [JETP Lett. {\bf 9}, 87 (1969)]
\bibitem{Gorkov} GL theory is derived from the microscopic
Bogoliubov - de Gennes equations assuming that spatial variations
in $\Delta(T)$ take place on length scales much larger than $\xi_0$.
See L. P. Gor'kov, Zh. Eksp. Teor. Fiz. {\bf 36}, 1918 (1959)
(Sov. Phys. JETP {\bf 9}, 1364 (1960)).
This condition may not be fulfilled for a typical mesoscopic sample
where geometrical boundaries introduce strong variations on $\Delta(T)$
on length scales of the order or smaller than $\xi_0$.
\bibitem{Eilenberger} One should mention that there has been a large
number of works
(specially during the 70's) devoted to the study of SWL and based
on simplified semiclassical
versions of the complete Bogoliubov - de Gennes equations.
For a review on these works see ref. \cite{Likharev}.
\bibitem{de Gennes} P. G. de Gennes, {\it Superconductivity
of metals and alloys} (Benjamin, New York 1966).
\bibitem{jap} A. Furusaki and M. Tsukada, Solid State Commun. {\bf 78},
299 (1991).
\bibitem{Sols} F. Sols and J. Ferrer, Phys. Rev. B {\bf 49}, 15913 (1994).
\bibitem{Keldysh} L. V. Keldysh, Sov. Phys. JETP {\bf 20}, 1018 (1965).
\bibitem{Nambu} Y. Nambu, Phys. Rev. {\bf 117}, 648 (1960).
\bibitem{Kadanoff} L. P. Kadanoff and G. Baym in {\it Quantum
Statistical Mechanics} (Benjamin, New York, 1962).
\bibitem{Yeyati} See for instance,
A. Levy Yeyati, Phys.Rev.B {\bf 45}, 14189 (1992).
\bibitem{Arnold} G. B. Arnold, J. Low Temp. Phys. {\bf 59}, 143 (1985).
\bibitem{Kulik} O. Kulik and A. N. Omel'yanchuk, Fiz. Nisk. Temp. {\bf 3},
945 (1977); {\bf 4}, 296 (1978) [Sov. J. Low Temp. Phys. {\bf 3}, 459
(1977); {\bf 4}, 142 (1978)].
\bibitem{Ambegaokar} V. Ambegaokar and A. Baratoff, Phys.Rev.Lett.
{\bf 10}, 486 (1963); ibid. Phys.Rev.Lett. {\bf 11}, 104 (1963).
\bibitem{Bagwell2} P. F. Bagwell, Phys. Rev. B {\bf 49}, 6841 (1994).
\bibitem{Arnold2} M. J. DeWeert and G. B. Arnold, Phys. Rev. Lett. {\bf 55},
1522 (1985); M. J. DeWeert and G. B. Arnold, Phys. Rev. B {\bf 39}, 11307
(1989).
\bibitem{Pablo} P. L. Pernas, A. Mart\'{\i}n-Rodero and F. Flores,
 Phys. Rev. B {\bf 41}, 8553 (1990).
\bibitem{Butt2} M. Buttiker, Phys. Rev. B {\bf 40}, 3409 (1989)
\bibitem{Pasta} J. L. D' Amato and H. Pastawski, Phys. Rev. B
{\bf 41}, 7411 (1990).
\bibitem{Langer} J. S. Langer and V. Ambegaokar, Phys. Rev. {\bf 164},
498 (1967).
\bibitem{Vander} D. Vanderbilt and S. G. Louie, Phys. Rev. B {\bf 30},
6118 (1984).
\end{references}
\end{document}